 \documentclass[final,3p,times,twocolumn]{elsarticle}

\usepackage{amssymb}

\biboptions{sort&compress}

\journal{NIM B336(2014)37-44}

\begin{document}

\begin{frontmatter}



\title{New data on activation cross section for deuteron induced reactions on ytterbium up to 50 MeV}


\author[1]{F. T\'ark\'anyi}
\author[1]{F. Ditr\'oi\corref{*}}
\author[1]{S. Tak\'acs}
\author[2]{A. Hermanne}
\author[4]{A.V. Ignatyuk}
\cortext[*]{Corresponding author: ditroi@atomki.hu}

\address[1]{Institute for Nuclear Research, Hungarian Academy of Sciences (ATOMKI),  Debrecen, Hungary}
\address[2]{Cyclotron Laboratory, Vrije Universiteit Brussel (VUB), Brussels, Belgium}
\address[4]{Institute of Physics and Power Engineering (IPPE), Obninsk, Russia}

\begin{abstract}
Activation cross sections of deuteron induced reactions on ytterbium for production of $^{177g,173,172,171,170,169,167}$Lu, $^{177,175,169}$Yb and $^{173,168,167,165}$Tm were extended up to 50 MeV deuteron energy. The new data are in acceptable agreement with the earlier experimental data in the overlapping energy region. The experimental data are compared with the predictions of the ALICE-D, EMPIRE-D and TALYS 1.4 (TENDL-2013 on-line library results) codes.
\end{abstract}


\end{frontmatter}


\section{Introduction}
\label{1}
To meet the requirements of improving the reliability of available experimental and theoretical cross section data, we started to establish an experimental database some years ago, by performing new experiments and making a systematical survey of published deuteron induced activation cross-sections up to 50 MeV \cite{1}. In an earlier publication we presented the activation cross-sections of longer-lived products of deuteron induced nuclear reactions on ytterbium up to 40 MeV, obtained in irradiations at the Sendai cyclotron \cite{2}. The earlier data of Nichols et al. \cite{3}, Hermanne et al., \cite{4}, Manenti et al. \cite{5} and Dmitriev et al. \cite{6} were discussed in more detail at energies below 30 MeV in this publication. Since that time only one new work was published by Manenti et al. on physical optimization of production of high speciﬁc activity $^{177g}$Lu by deuteron irradiation \cite{7}.  We have had now the possibility to extend the energy range up to 50 MeV deuteron energy and to investigate some shorter-lived reaction products (in the 40 MeV experiments the first $\gamma$-spectra could only be measured one day after end of bombardment). To avoid repetition of the content of our previous paper \cite{2} we describe the experiment, the results and the theoretical comparisons only in summary form.

\section{Experiment and data evaluation}
\label{2}
For measurements, the well-known stacked foil irradiation technique and high resolution $\gamma$-spectrometry were used. Yb metal foils, interleaved with Al foils for monitoring beam characteristics were stacked and irradiated at the UCL (LLN) cyclotron. The Yb foils were irradiated together with Nd foils. The report on activation cross section data on Nd is in progress \cite{8}.
The main experimental parameters and methods of data evaluation are summarized in Table 1. The complete excitation function was measured for the $^{nat}$Al(d,x)$^{24}$Na monitor reactions, allowing to control the beam intensity and the energy by comparison with recommended data \cite{9}, and are shown in Fig. 1 in our earlier submitted paper on nuclear reactions on simultaneously irradiated Nd \cite{7}. The decay characteristics of the investigated reaction products and the possibly contributing reactions in the energy region studied are summarized in Table 2.
It should be mentioned that in a few cases we could not find independent $\gamma$-lines to assess the produced activity of the investigated radioisotopes. In these cases the contributions of the overlapping $\gamma$-lines from the decay of the other nuclides were subtracted.

\begin{table*}[t]
\tiny
\caption{Main experimental parameters and  main parameters and methods of the data evaluation}
\centering
\begin{center}
\begin{tabular}{|p{1.3in}|p{1.4in}|p{1.5in}|p{1.5in}|} \hline 
\multicolumn{2}{|p{1in}|}{\textbf{Main experimental parameters}} & \multicolumn{2}{|p{3.0in}|}{\textbf{Methods of the data evaluation}} \\ \hline 
Incident particle & Deuteron (LLN) & Gamma spectra evaluation & Genie 2000\cite{10}, Forgamma \cite{11} \\ \hline 
Method  & Stacked foil & Determination of beam intensity & Faraday cup (preliminary)\newline Fitted monitor reaction (final) \cite{12} \\ \hline 
Target composition & $^{nat}$Nd (100 $\mu$m)-target\newline $^{nat}$Yb(22.88 $\mu$m)-target\newline  Al \textbf{(}49.06 $\mu$m)-monitor\newline (repeated 19 times) \newline Interleaved with Al (156.6 $\mu$m, 103.43 $\mu$m, 49.06 $\mu$m)-energy degraders\newline \newline  & Determination of beam intensity & Faraday cup (preliminary)\newline Fitted monitor reaction (final)\cite{9, 12}  \\ \hline 
Number of Yb  target foils & 19 & Decay data (see Table 2) & NUDAT 2.6 \cite{13} \\ \hline 
Accelerator & Cyclone110 cyclotron of Université Catholique in Louvain la Neuve (LLN) Belgium & Reaction Q-values(see Table 2) & Q-value calculator \cite{14} \\ \hline 
Primary energy & 50 MeV & Determination of beam energy & Andersen (preliminary\cite{15}\newline Fitted monitor reaction (final)\newline  \cite{9}  \\ \hline 
Covered energy range & 48.2-12.9 MeV & Uncertainty of energy & Cumulative effects of possible uncertainties\newline (primary energy, target thickness, energy straggling,  correction to monitor reaction) \\ \hline 
Irradiation time & 60  min & Cross sections & Isotopic and elemental cross sections \\ \hline 
Beam current & 92 nA & Uncertainty of cross sections & Sum in quadrature of all individual contributions: beam current (7\%), beam-loss corrections (max. of 1.5\%), target thickness (1\%), \newline detector efficiency (5\%),  \newline photo peak area determination  and counting statistics (1-20 \cite{16} \\ \hline 
Monitor reaction [recommended values]  & $^{27}$Al(d,x)$^{24}$Na  reaction \cite{9} \newline (re-measured over the whole energy range) &  &  \\ \hline 
Monitor target and thickness & ${}^{Cityplacenat}$Al, 49.06 mm &  &  \\ \hline 
detector & HPGe &  &  \\ \hline 
Chemical separation & no &  &  \\ \hline 
$\gamma$-spectra measurements & 3 series &  &  \\ \hline 
Cooling times\newline (and corresponding target-detector distances) & 2.1-5.5 h (25 cm)\newline 23.7-30.7 h (15 cm)\newline 36.9-432.2 h (5cm)\newline  &  &  \\ \hline 
\end{tabular}
\end{center}

\end{table*}

\begin{table*}[t]
\tiny
\caption{Decay characteristics of the investigated reaction products and the contributing reactions}
\centering
\begin{center}
\begin{tabular}{|p{0.3in}|p{0.3in}|p{0.3in}|p{0.2in}|p{0.4in}|p{0.3in}|} \hline 
Nuclide & Half-life & E$_{\gamma}$(keV) & I$_{\gamma }$(\%) & Contributing reaction & Q-value\newline (MeV) \\ \hline 
$^{177g}$Lu\newline $\varepsilon $: 100~\% & 6.647 d & 112.9498\newline 208.3662 & 6.17\newline 10.36 & ${}^{176}$Yb(d,n)\newline ${}^{177}$Yb decay & 3.9 \\ \hline 
${}^{173}$Lu\newline $\varepsilon $: 100~\% & 1.37 a & 78.63\newline 100.724\newline 272.105 & 11.9\newline 5.24\newline 21.2 & ${}^{172}$Yb(d,n)\newline ${}^{173}$Yb(d,2n)\newline ${}^{174}$Yb(d,3n)\newline ${}^{176}$Yb(d,5n) & 2.7\newline -3.7\newline -11.1\newline -23.9 \\ \hline 
${}^{172g}$Lu\newline $\varepsilon $: 100~\% & 6.70 d & 78.7426\newline 181.525\newline 810.064\newline 900.724\newline 912.079\newline 1093.63 & 10.6 \newline 20.6 \newline 16.6 \newline 29.8\newline 15.3\newline 63  & ${}^{171}$Yb(d,n)\newline ${}^{172}$Yb(d,2n)\newline ${}^{173}$Yb(d,3n)\newline ${}^{174}$Yb(d,4n)\newline ${}^{176}$Yb(d,6n) & 2.5\newline -5.5\newline -11.9\newline -19.4\newline -32.0 \\ \hline 
${}^{171g}$Lu\newline $\varepsilon $: 100~\% & 8.24 d & 667.422\newline 739.793\newline 780.711\newline 839.961 & 11.1 \newline 47.9 \newline 4.37\newline 3.05 & ${}^{170}$Yb(d,n)\newline ${}^{171}$Yb(d,2n)\newline ${}^{172}$Yb(d,3n)\newline ${}^{173}$Yb(d,4n)\newline ${}^{174}$Yb(d,5n)\newline ${}^{176}$Yb(d,7n) & 2.1\newline -4.5\newline -12.5\newline -18.9\newline -26.3\newline -39.0 \\ \hline 
${}^{170}$Lu\newline $\varepsilon $: 100~\% & 2.012 d & 84.262\newline 193.13\newline 572.20\newline 985.10\newline 1054.28\newline 1138.65\newline 1280.25\newline 1341.20\newline 1364.60   & 8.7 \newline 2.07 \newline 1.25 \newline 5.4  \newline 4.60\newline 3.49 \newline 7.9 \newline 3.15\newline 4.47 & ${}^{170}$Yb(d,2n)\newline ${}^{171}$Yb(d,3n)\newline ${}^{172}$Yb(d,4n)\newline ${}^{173}$Yb(d,5n)\newline ${}^{174}$Yb(d,6n)\newline ${}^{176}$Yb(d,8n) & -6.5\newline -13.1\newline -21.1\newline -27.5\newline -34.9\newline -47.6 \\ \hline 
${}^{169}$Lu\newline $\varepsilon $: 100~\% & 34.06 h & 191.217\newline 960.622 & 18.7 \newline 21.2 \newline  & ${}^{168}$Yb(d,n)\newline ${}^{170}$Yb(d,3n)\newline ${}^{171}$Yb(d,4n)\newline ${}^{172}$Yb(d,5n)\newline ${}^{173}$Yb(d,6n)\newline ${}^{174}$Yb(d,7n)\newline ${}^{176}$Yb(d,9n) & 1.6\newline -13.8\newline -20.4\newline -28.4\newline -34.8\newline -42.2\newline -54. \\ \hline 
${}^{167}$Lu\newline $\varepsilon $: 100~\% & 51.5 min & 178.87\newline 213.20\newline 239.22\newline 401.17\newline 1267.26 & 2.5 \newline 3.33\newline 7.7 \newline 3.17\newline 3.87    & ${}^{168}$Yb(d,3n)\newline ${}^{170}$Yb(d,5n)\newline ${}^{171}$Yb(d,6n)\newline ${}^{172}$Yb(d,7n)\newline ${}^{173}$Yb(d,8n)\newline ${}^{174}$Yb(d,9n)\newline ${}^{176}$Yb(d,11n) & -15.2\newline -30.5\newline -37.1\newline -45.1\newline -51.5\newline -59.0 \\ \hline 
${}^{177}$Yb\newline $\beta $${}^{-}$: 100~\% & 1.911 h & 150.3 & 20.5 & ${}^{176}$Yb(d,p) & 3.3 \\ \hline 
${}^{175}$Yb\newline $\beta $${}^{-}$: 100~\% & 4.185 d & 113.805\newline 282.522\newline 396.329 & 3.87\newline 6.13 \newline 13.2 & ${}^{174}$Yb(d,p)\newline ${}^{176}$Yb(d,p2n)\newline ${}^{175}$Tm decay & 3.6\newline -9.1 \\ \hline 
${}^{1}$${}^{69}$Yb\newline $\varepsilon $: 100~\%\newline  & 32.018 d & 109.77924\newline 130.52293\newline 177.21307\newline 197.95675\newline 307.52\newline 307.73586 & 17.39\newline 11.38\newline 22.28\newline 35.93\newline 0.3\newline 10.05\newline \newline  & ${}^{168}$Yb(d,p)\newline ${}^{170}$Yb(d,p2n)\newline ${}^{171}$Yb(d,p3n)\newline ${}^{172}$Yb(d,p4n)\newline ${}^{173}$Yb(d,p5n)\newline ${}^{174}$Yb(d,p6n)\newline ${}^{176}$Yb(d,p8n)\newline ${}^{169}$Lu decay & 4.6\newline -10.7\newline -17.3\newline -25.3\newline -31.7\newline -39.2\newline -51.9 \\ \hline 
${}^{173}$Tm${}^{ }$\newline $\beta $${}^{-}$: 100~\% & 8.24 h & 398.9\newline 461.4 & 87.9 \newline 6.9  & ${}^{173}$Yb(d,2p)\newline ${}^{174}$Yb(d,2pn)\newline ${}^{176}$Yb(d,2p3n) & -2.7\newline -10.2\newline -22.9 \\ \hline 
${}^{172}$Tm${}^{   }$\newline $\beta $${}^{-}$: 100~\% & ~63.6 h & 78.750 \newline 181.520\textit{\newline }1093.59\newline 1387.093 & 6.5\newline 2.8\newline 6.0 \newline 5.6  & ${}^{172}$Yb(d,2p)\newline ${}^{173}$Yb(d,2pn)\newline ${}^{174}$Yb(p,2p2n)\newline ${}^{176}$Yb(p,2p4n) & -3.3\newline -9.7\newline -17.2\newline -29.8 \\ \hline 
${}^{168}$Tm${}^{ }$\newline ${}^{ }$$\varepsilon $: 99.99~\%\newline $\beta $${}^{-}$: 0.01~\%\newline  & 93.1 d & 79.804\newline 184.295\newline 198.251\newline 447.515\newline 720.392\newline 741.355\newline 815.989\newline 821.162 & 10.95\newline 18.55\newline 54.49\newline 23.98\newline 12.207\newline 12.81 \newline 50.95 \newline 11.99 & ${}^{168}$Yb(d,2p)\newline ${}^{170}$Yb(d,2p2n)\newline ${}^{171}$Yb(d,2p3n)\newline ${}^{172}$Yb(d,2p4n)\newline ${}^{173}$Yb(d,2p5n)\newline ${}^{174}$Yb(d,2p6n)\newline ${}^{176}$Yb(d,2p8n) & -1.7\newline -17.0\newline -23.7\newline -31.7\newline -38.0\newline -45.5\newline -58.2 \\ \hline 
${}^{167}$Tm${}^{  }$\newline $\varepsilon $: 100~\% & 9.25 d & 207.801\newline 531.54 & 42 \newline 1.61 & ${}^{168}$Yb(d,2pn)\newline ${}^{170}$Yb(d,2p3n)\newline ${}^{171}$Yb(d,2p4n)\newline ${}^{172}$Yb(d,2p5n)\newline ${}^{173}$Yb(d,2p6n)\newline ${}^{174}$Yb(d,2p7n)\newline ${}^{176}$Yb(d,2p9n)\newline ${}^{167}$Yb decay & -8.5\newline -23.9\newline -30.5\newline -38.5\newline -44.9\newline -52.3\newline -65.0\newline  \\ \hline 
${}^{165}$Tm${}^{ }$\newline ${}^{ }$$\varepsilon $: 100~\% & 30.06 h & 242.917\newline 296.49\newline 297.369\newline 460.263 & 35.5\newline 3.88\newline 12.7 \newline 4.12 & ${}^{168}$Yb(d,2p3n)\newline ${}^{170}$Yb(d,2p5n)\newline $^{171}$Yb(d,2p6n)\newline ${}^{172}$Yb(d,2p7n)\newline ${}^{173}$Yb(d,2p8n)\newline ${}^{174}$Yb(d,2p9n)\newline ${}^{176}$Yb(d,2p11n)\newline ${}^{165}$Yb decay & -24.3\newline -39.6\newline -46.2\newline -54.3\newline -60.6\newline -68.1 \\ \hline 
\end{tabular}

\end{center}

\begin{flushleft}
\tiny{
\noindent *When complex particles are emitted instead of individual protons and neutrons the Q-values have to be decreased by the respective binding energies of the compound particles: np-d, +2.2 MeV; 2np-t, +8.48 MeV; 2p2n-a, 28.30 MeV.

\noindent **Abundance of isotopes in natural Yb (\%): $^{168}$Yb-0.13, $^{170}$Yb-3.05, $^{171}$Yb-14.3, $^{172}$Yb-21.9, $^{173}$Yb-16.12, $^{174}$Yb-31.8, $^{176}$Yb-12.7.
\noindent ***The Q-values refer to formation of the ground state and were obtained from \cite{14}
 }

\end{flushleft}

\end{table*}

\section{Comparison with the results of model codes}
\label{3}
In our previous work we made calculations for the investigated reactions using the modified model codes ALICE-IPPE \cite{17} and EMPIRE-II \cite{18}. In the used modified code versions ALICE-IPPE-D and EMPIRE-D, the direct (d,p) channel is increased strongly \cite{19, 20}. Here we repeat these results for comparison above 40 MeV too. The new experimental data are also compared with the cross section data reported in the TALYS 1.4 based \cite{21} TENDL-2013 data libraries \cite{22}.

\section{Results}
\label{4}
The cross-sections for all reactions investigated are shown in Figs. 1–14 and the numerical values are shown in Tables 3-4. The contributing reactions can be found in Table 2. The reactions were discussed in detail in our previous work \cite{2}. The agreement (or disagreement) of the new data with the previous experimental data and with the model results are shown in the corresponding figures and discussed below.
The new results are in acceptable agreement with the previous data in most cases. As in \cite{2} we deduced already integral yields for production of the investigated reaction products up to 40 MeV, we did not include in this paper a new figure  extended to 50 MeV.

\subsection{$^{177g}$Lu (cum)}
\label{4.1}

The cumulative production of $^{177g}$Lu (6.71 d half-life) following the total decay of parent $^{177}$Yb (T$_{1/2}$ = 1.9 h) was detected. It practically contains no contribution from the internal decay of $^{177m}$Lu (160.4 d, IT 21.7\%) (long-lived, low formation cross-section). Only two cross section points were reliably assessed, near the maximum (Fig. 1). The new data are in good agreement with our earlier results, but somewhat higher than the results of Manenti \cite{5} and Hermanne \cite{4}. The nuclear model codes, especially the TENDL-2013 give lower values in this energy region.

\begin{figure}
\includegraphics[scale=0.3]{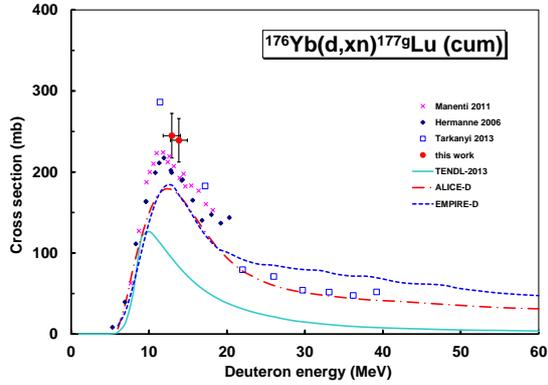}
\caption{Excitation function of the $^{176}$Yb(d,x)$^{177g}$Lu reaction}
\end{figure}

\subsection{$^{173}$Lu}
\label{4.2}
The production of $^{173}$Lu (1.37 a) arises from $^{nat}$Yb(d,xn)$^{173}$Lu reactions on different stable Yb isotopes (Fig. 2). The new data are in good agreement with our earlier results \cite{2} and also with those of Manenti \cite{5} and Hermanne \cite{4}. The values of Nichols \cite{3} do not even reproduce the first maximum. The nuclear model codes follow the shape of the experimental curve and also the values, but it is difficult to judge, which of them gives the best approximation.

\begin{figure}
\includegraphics[scale=0.3]{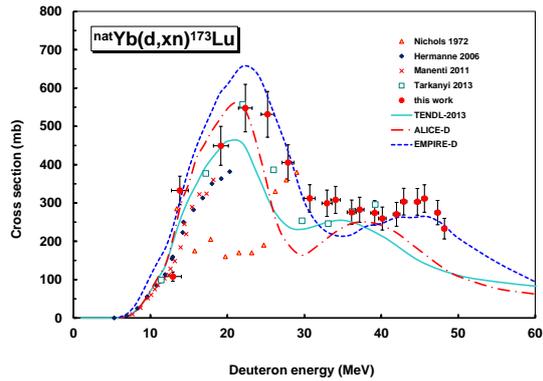}
\caption{Excitation function of the  $^{nat}$Yb(d,xn)$^{173}$Lu reaction}
\end{figure}

\subsection{$^{172g}$Lu (m+)}
\label{4.3}
The cross sections contain the complete contribution of the decay of short-lived isomeric state (3.7 min) (Fig. 3). Our new data are in good agreement with the earlier experimental results. The best approximation is given by the TALYS (TENDL-2013) nuclear reaction model code, while both other codes overestimate the maximum and also the maximum energy.

\begin{figure}
\includegraphics[scale=0.3]{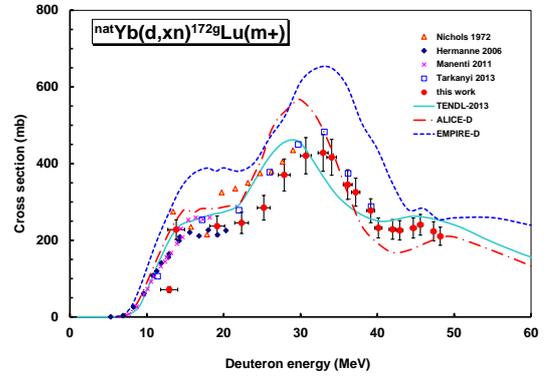}
\caption{Excitation function of the  $^{nat}$Yb(d,xn)$^{172mg}$Lu reaction}
\end{figure}

\subsection{$^{171g}$Lu (m+)}
\label{4.4}
The cumulative production of 171Lu (8.24 d) includes the complete decay through isomeric transition of the short-lived (79 s) isomeric state (Fig. 4). The new data are in good agreement with our former results, with the experimental results of Hermanne \cite{4} and Manenti \cite{5}, but give somewhat lower values than our previous results between 20 and 40 MeV. The best estimate is given by the TENDL-2013 prediction. 

\begin{figure}
\includegraphics[scale=0.3]{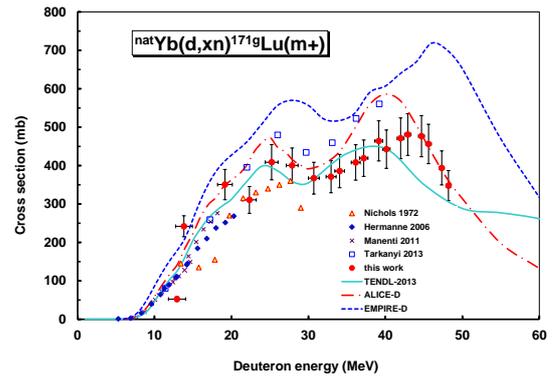}
\caption{Excitation function of the  $^{nat}$Yb(d,xn)$^{171mg}$Lu reaction}
\end{figure}

\subsection{$^{170}$Lu}
\label{4.5}
No parent contribution exists for the formation of $^{170}$Lu (no isomeric state), so the presented results are direct cross-sections resulting from (d,xn) reactions (Fig. 5). Our new data are in good agreement with the previous experimental results, except the first local maximum, where our previous results were slightly higher. The best computational approximation is given by the TENDL-2013 library again.

\begin{figure}
\includegraphics[scale=0.3]{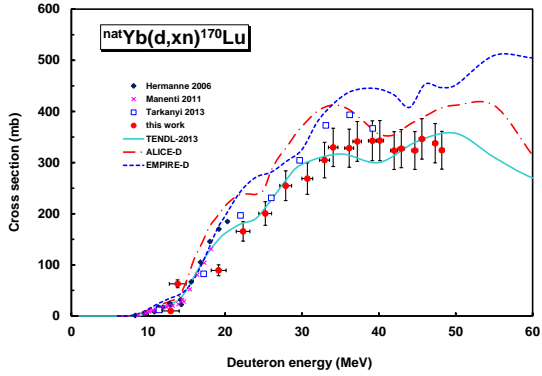}
\caption{Excitation function of the  $^{nat}$Yb(d,xn)$^{170}$Lu reaction}
\end{figure}

\subsection{$^{169}$Lu}
\label{4.6}
$^{169}$Lu (32.018 d) is produced directly via the $^{nat}$Yb(d,xn) reactions (Fig. 6). The new data are in good agreement with our previous experimental results. The best model approach is given by the TENDL-2013 library again.

\begin{figure}
\includegraphics[scale=0.3]{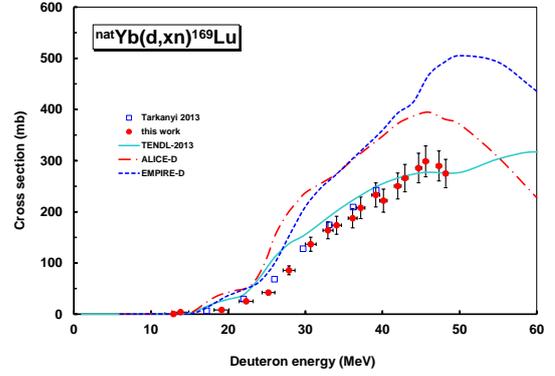}
\caption{Excitation function of the  $^{nat}$Yb(d,xn)$^{169}$Lu reaction}
\end{figure}

\subsection{$^{167}$Lu}
\label{4.7}
No earlier experimental data were found for the formation of $^{167}$Lu through the $^{nat}$Yb(d,xn)$^{167}$Lu reactions. Our new data for production of $^{167}$Lu (51.5 min) are presented in Fig. 7. Our new experimental data are below all predictions of the nuclear reaction model codes.

\begin{figure}
\includegraphics[scale=0.3]{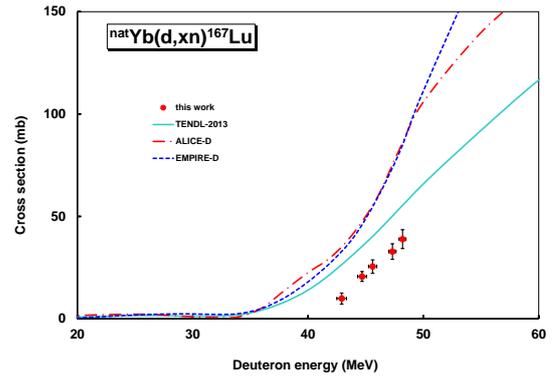}
\caption{Excitation function of the  $^{nat}$Yb(d,xn)$^{167}$Lu reaction}
\end{figure}

\subsection{$^{177}$Yb}
\label{4.8}
In our previous measurement up to 40 MeV \cite{2} we could not identify the $\gamma$-lines of $^{177}$Yb (1.911 h) in our spectra due to the long cooling time. As in these experiments measurements could be started 2 h after EOB, a statistically significant signal for the independent 150.3 keV line of $^{177}$Yb could be identified. The new results are shown in Fig. 8, and are in good agreement with \cite{4} and \cite{5}. $^{177}$Yb can only be produced via the $^{176}$Yb(d,p)$^{177}$Yb reaction. All nuclear reaction model codes underestimate the maximum value, the best estimate is given by the ALICE-D code (due to its improved (d,p) capability).

\begin{figure}
\includegraphics[scale=0.3]{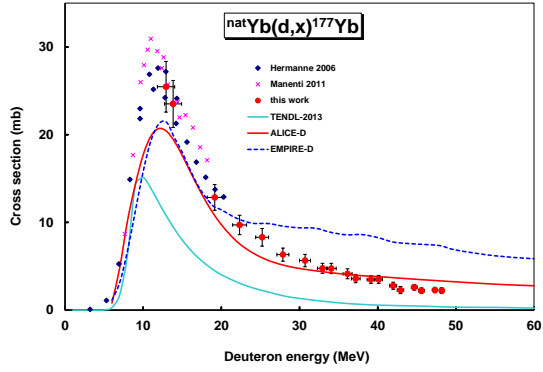}
\caption{Excitation function of the  $^{176}$Yb(d,p)$^{177}$Yb reaction }
\end{figure}

\subsection{$^{175}$Yb (cum)}
\label{4.9}
The cumulative production of $^{175}$Yb (4.185 d) (via direct (d,pxn) reactions and from $\beta^-$-decay of $^{175}$Tm (15.2 min) )  is shown in Fig. 9.  Our new data support the earlier data from our group [2, 5] in the lower energy region due to the $^{176}$Yb(d,2pn) reaction. All nuclear reaction codes underestimate the experimental values, only the first maximum energy is given correctly by the ALICE-D and EMPIRE-D codes. The second broad maximum is only predicted by the TENDL-2013 library.

\begin{figure}
\includegraphics[scale=0.3]{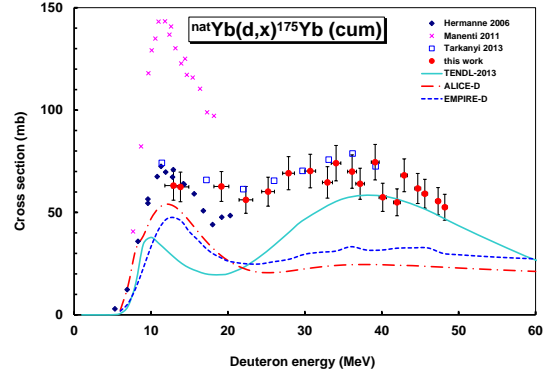}
\caption{Excitation function of the  $^{nat}$Yb(d,x)$^{175}$Yb process}
\end{figure}

\subsection{$^{169}$Yb (cum)}
\label{4.10}
The measured cumulative activation cross-sections of $^{169}$Yb (T$_{1/2}$ = 32.018 d) are shown in Fig. 10. This radioisotope is obtained through direct production via $^{nat}$Yb(d,pxn) reactions and from the  decay of the shorter-lived parent $^{169}$Lu (34.06 h). Our new data are in good agreement with our previous results in the overlapping energy range and also with the other experiment in the low energy region. Now, all the nuclear reaction model codes give similar results in the measured energy region. The TENDL-2013 prediction seems to be the closest approximation, but it does not predict the expected maximum below 50 MeV.

\begin{figure}
\includegraphics[scale=0.3]{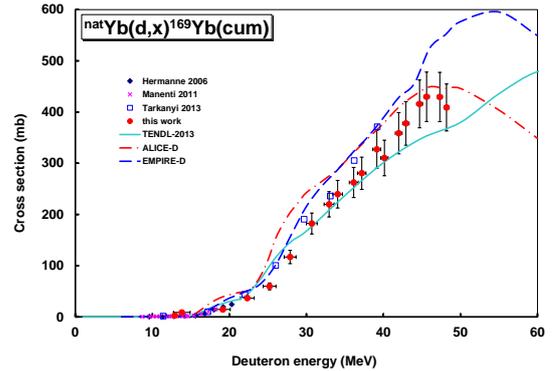}
\caption{Excitation function of the  $^{nat}$Yb(d,x)$^{169}$Yb reaction}
\end{figure}

\subsection{$^{173}$Tm}
\label{4.11}
The cross sections of directly produced $^{173}$Tm (8.24 h) are presented on Fig. 11. In spite of the quite strong scattering, our new experimental data support our previous measurement. EMPIRE-D and ALICE-D give acceptable estimates below 27 MeV, reproducing the first local maximum too, while TALYS completely fails in this case.

\begin{figure}
\includegraphics[scale=0.3]{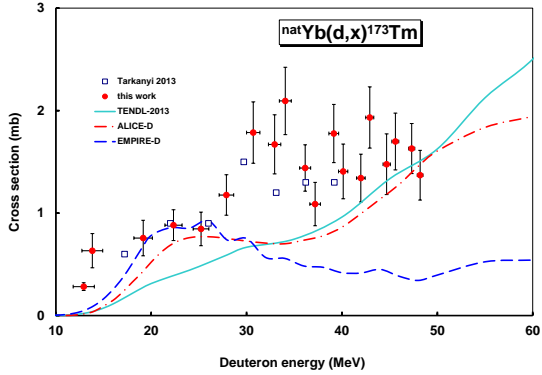}
\caption{Excitation function of the  $^{nat}$Yb(d,x)$^{173}$Tm reaction}
\end{figure}

\subsection{$^{168}$Tm}
\label{4.12}
The measured excitation function for direct production of $^{168}$Tm (93.1 d) is shown in Fig. 12. Because of the low statistics the data are scattered again, but the good agreement with our previous results is obvious. All nuclear reaction model codes, especially TALYS, underestimate the experimental values.

\begin{figure}
\includegraphics[scale=0.3]{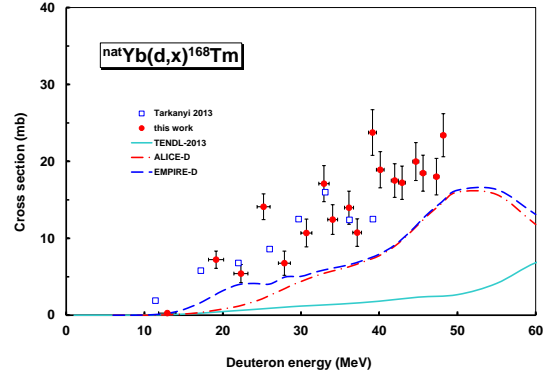}
\caption{Excitation function of the  $^{nat}$Yb(d,x)$^{168}$Tm reaction}
\end{figure}

\subsection{$^{167}$Tm (cum)}
\label{4.13}
The excitation function for cumulative production of $^{167}$Tm (9.25 d) (direct reactions and from the decay of short-lived parent $^{167}$Yb (17.5 min)) are sown in Fig. 13. The best prediction is provided by the TENDL-2013 library above 35 MeV, while under 35 MeV all the three codes show approximately the same results.

\begin{figure}
\includegraphics[scale=0.3]{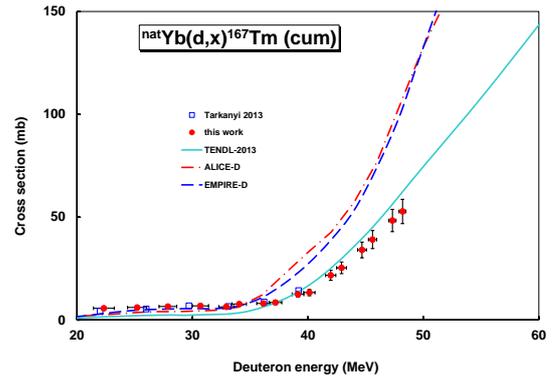}
\caption{Excitation function of the  $^{nat}$Yb(d,x)$^{167}$Tm reaction}
\end{figure}

\subsection{$^{165}$Tm (cum)}
\label{4.14}
The measured excitation function for $^{165}$Tm (30.06 h) is shown in Fig. 14. It was produced both directly through $^{nat}$Yb(d,pxn) reactions and indirectly from decay of short-lived parent $^{165}$Yb (9.9 min). The overlap with our previous results is acceptable. The best prediction is given by the ALICE-D code in this case.

\begin{figure}
\includegraphics[scale=0.3]{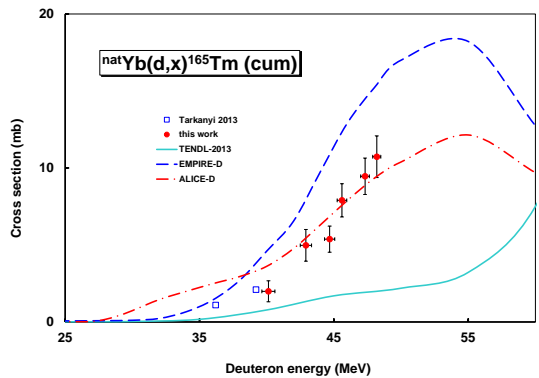}
\caption{Excitation function of the  $^{nat}$Yb(d,x)$^{165}$Tm reaction}
\end{figure}

\begin{table*}[t]
\tiny
\caption{Measured cross-sections of the $^{nat}$Yb(d,xn)$^{177,173,172mg,171mg,170,169}$Lu reactions}
\centering
\begin{center}
\begin{tabular}{|p{0.3in}|p{0.3in}|p{0.3in}|p{0.2in}|p{0.3in}|p{0.2in}|p{0.3in}|p{0.2in}|p{0.3in}|p{0.2in}|p{0.3in}|p{0.2in}|p{0.3in}|p{0.2in}|} \hline 
\multicolumn{2}{|c|}{Energy E$\pm\Delta$E\newline    (MeV)} & \multicolumn{12}{|c|}{Cross section $\sigma\pm\Delta\sigma$ (mb)} \\ \hline 
\multicolumn{2}{|c|}{} & \multicolumn{2}{|c|}{$^{177}$Lu} & \multicolumn{2}{|c|}{$^{173}$Lu} & \multicolumn{2}{|c|}{$^{172g}$Lu} & \multicolumn{2}{|c|}{$^{171g}$Lu} & \multicolumn{2}{|c|}{${}^{170}$Lu} & \multicolumn{2}{|c|}{${}^{16}$${}^{9}$Lu         } \\ \hline 
48.2 & 0.3 & ~ &  & 233.1 & 26.8 & 210.5 & 23.8 & 347.8 & 39.2 & 324.2 & 37.0 & 274.9 & 27.8 \\ \hline 
47.3 & 0.3 & ~ &  & 274.9 & 31.8 & 223.4 & 25.2 & 393.9 & 44.4 & 337.8 & 38.5 & 289.6 & 29.3 \\ \hline 
45.6 & 0.4 & ~ &  & 311.6 & 35.8 & 240.4 & 27.1 & 456.4 & 51.4 & 346.3 & 39.5 & 298.6 & 30.2 \\ \hline 
44.7 & 0.4 & ~ &  & 303.0 & 34.8 & 231.9 & 26.2 & 476.6 & 53.7 & 323.6 & 37.0 & 285.6 & 28.9 \\ \hline 
42.9 & 0.4 & ~ &  & 303.7 & 34.8 & 226.0 & 25.5 & 480.9 & 54.1 & 327.2 & 37.3 & 265.9 & 26.9 \\ \hline 
42.0 & 0.4 & ~ &  & 270.6 & 31.1 & 228.4 & 25.8 & 471.0 & 53.0 & 323.4 & 36.9 & 250.4 & 25.3 \\ \hline 
40.1 & 0.5 & ~ &  & 259.6 & 30.0 & 232.2 & 26.2 & 442.8 & 49.9 & 342.9 & 39.2 & 221.8 & 22.4 \\ \hline 
39.1 & 0.5 & ~ &  & 274.2 & 31.8 & 277.3 & 31.2 & 464.2 & 52.2 & 342.6 & 39.0 & 232.9 & 23.6 \\ \hline 
37.2 & 0.6 & ~ &  & 282.5 & 32.7 & 325.4 & 36.7 & 419.3 & 47.2 & 341.3 & 38.9 & 207.8 & 21.0 \\ \hline 
36.1 & 0.6 & ~ &  & 275.9 & 31.9 & 345.3 & 38.9 & 408.3 & 46.0 & 328.2 & 37.3 & 187.5 & 19.0 \\ \hline 
34.1 & 0.6 & ~ &  & 308.0 & 35.3 & 416.4 & 46.9 & 385.8 & 43.5 & 329.7 & 37.6 & 173.8 & 17.6 \\ \hline 
32.9 & 0.7 & ~ &  & 299.4 & 34.4 & 428.1 & 48.2 & 371.0 & 41.8 & 304.8 & 34.9 & 163.9 & 16.6 \\ \hline 
30.7 & 0.7 & ~ &  & 312.0 & 35.8 & 420.6 & 47.3 & 367.3 & 41.4 & 268.6 & 30.7 & 136.7 & 13.8 \\ \hline 
27.9 & 0.8 & ~ &  & 405.4 & 46.3 & 370.5 & 41.7 & 400.5 & 45.1 & 254.6 & 29.1 & 85.6 & 8.7 \\ \hline 
25.2 & 0.8 & ~ &  & 531.5 & 59.8 & 284.8 & 32.1 & 408.9 & 46.1 & 200.5 & 23.1 & 42.2 & 4.3 \\ \hline 
22.3 & 0.9 & ~ &  & 547.7 & 62.1 & 245.5 & 27.7 & 310.7 & 35.0 & 165.4 & 19.0 & 25.2 & 2.6 \\ \hline 
19.2 & 1.0 & ~ &  & 449.3 & 50.8 & 236.9 & 26.7 & 350.6 & 39.5 & 89.3 & 10.7 & 8.3 & 0.9 \\ \hline 
13.8 & 1.1 & 239 & 27 & 332.6 & 37.4 & 227.9 & 25.7 & 242.0 & 27.3 & 62.9 & 7.7 & 4.0 & 0.5 \\ \hline 
12.9 & 1.1 & 245 & 28 & 108.0 & 12.2 & 71.3 & 8.0 & 52.6 & 5.9 & 9.9 & 1.1 & 0.3 & 0.04 \\ \hline 
\end{tabular}
\end{center}
\end{table*}

\begin{table*}[t]
\tiny
\caption{Measured cross-sections of the $^{nat}$Yb(d,x)$^{177,175,169}$Yb and $^{nat}$Yb(d,x)$^{173,168,167,165}$Tm}
\centering
\begin{center}
\begin{tabular}{|p{0.3in}|p{0.2in}|p{0.3in}|p{0.2in}|p{0.3in}|p{0.2in}|p{0.3in}|p{0.2in}|p{0.3in}|p{0.2in}|p{0.3in}|p{0.2in}|p{0.3in}|p{0.2in}|p{0.3in}|p{0.2in}|p{0.2in}|} \hline 
\multicolumn{2}{|c|}{Energy E$\pm\Delta$E\newline    (MeV)} &  & \multicolumn{12}{|c|}{Cross section $\sigma\pm\Delta\sigma$ (mb)} &  \\ \hline 
\multicolumn{2}{|c|}{} & \multicolumn{2}{|c|}{$^{177}$Yb} & \multicolumn{2}{|c|}{$^{175}$Yb} & \multicolumn{2}{|c|}{${}^{169}$Yb} & \multicolumn{2}{|c|}{$^{173}$Tm} & \multicolumn{2}{|c|}{$^{168}$Tm} &  & \multicolumn{2}{|c|}{$^{167}$Tm} & $^{165}$Tm \\ \hline 
48.2 & 0.3 & 2.2 & 0.3 & 52.6 & 6.3 & 408.9 & 45.9 & 1.37 & 0.24 & 23.4 & 2.8 & 52.8 & 5.9 & 10.7 & 1.4 \\ \hline 
47.3 & 0.3 & 2.3 & 0.3 & 55.5 & 6.6 & 429.5 & 48.2 & 1.63 & 0.24 & 18.0 & 2.4 & 48.4 & 5.4 & 9.5 & 1.2 \\ \hline 
45.6 & 0.4 & 2.2 & 0.3 & 59.2 & 7.0 & 429.8 & 48.3 & 1.70 & 0.28 & 18.5 & 2.3 & 39.1 & 4.4 & 7.9 & 1.1 \\ \hline 
44.7 & 0.4 & 2.6 & 0.3 & 61.7 & 7.4 & 415.9 & 46.7 & 1.48 & 0.30 & 20.0 & 2.5 & 34.1 & 3.8 & 5.4 & 0.8 \\ \hline 
42.9 & 0.4 & 2.3 & 0.4 & 68.1 & 8.0 & 377.8 & 42.4 & 1.93 & 0.30 & 17.2 & 2.2 & 25.4 & 2.9 & 5.0 & 1.0 \\ \hline 
42.0 & 0.4 & 2.8 & 0.4 & 54.9 & 6.6 & 358.7 & 40.3 & 1.34 & 0.23 & 17.5 & 2.2 & 21.7 & 2.5 & ~ &  \\ \hline 
40.1 & 0.5 & 3.5 & 0.5 & 57.4 & 6.9 & 310.3 & 34.9 & 1.41 & 0.27 & 18.9 & 2.4 & 13.3 & 1.5 & 2.0 & 0.7 \\ \hline 
39.1 & 0.5 & 3.5 & 0.4 & 74.5 & 8.7 & 327.2 & 36.8 & 1.78 & 0.28 & 23.8 & 3.0 & 12.4 & 1.4 & ~ &  \\ \hline 
37.2 & 0.6 & 3.6 & 0.5 & 64.0 & 7.6 & 280.5 & 31.5 & 1.09 & 0.21 & 10.7 & 1.8 & 8.5 & 1.0 & ~ &  \\ \hline 
36.1 & 0.6 & 4.1 & 0.6 & 70.0 & 8.1 & 262.5 & 29.5 & 1.44 & 0.23 & 14.0 & 2.2 & 8.0 & 0.9 & ~ &  \\ \hline 
34.1 & 0.6 & 4.7 & 0.6 & 74.1 & 8.7 & 239.6 & 26.9 & 2.09 & 0.33 & 12.4 & 1.9 & 7.7 & 0.9 & ~ &  \\ \hline 
32.9 & 0.7 & 4.8 & 0.6 & 64.7 & 7.7 & 219.9 & 24.7 & 1.67 & 0.29 & 17.1 & 2.3 & 6.4 & 0.8 & ~ &  \\ \hline 
30.7 & 0.7 & 5.7 & 0.7 & 70.2 & 8.2 & 182.4 & 20.5 & 1.79 & 0.30 & 10.7 & 1.8 & 6.9 & 0.8 & ~ &  \\ \hline 
27.9 & 0.8 & 6.3 & 0.8 & 69.2 & 8.1 & 117.0 & 13.2 & 1.18 & 0.20 & 6.8 & 1.6 & 6.5 & 0.8 & ~ &  \\ \hline 
25.2 & 0.8 & 8.3 & 1.0 & 60.2 & 7.1 & 59.7 & 6.7 & 0.85 & 0.16 & 14.1 & 1.7 & 6.1 & 0.7 & ~ &  \\ \hline 
22.3 & 0.9 & 9.7 & 1.1 & 56.1 & 6.5 & 36.8 & 4.2 & 0.88 & 0.15 & 5.4 & 1.2 & 5.7 & 0.7 & ~ &  \\ \hline 
19.2 & 1.0 & 12.8 & 1.5 & 62.7 & 7.3 & 14.5 & 1.6 & 0.76 & 0.17 & 7.2 & 1.1 & ~ &  & ~ &  \\ \hline 
13.8 & 1.1 & 23.5 & 2.7 & 62.5 & 7.3 & 8.7 & 0.98 & 0.63 & 0.17 & ~ &  & ~ &  & ~ &  \\ \hline 
12.9 & 1.1 & 25.5 & 2.9 & 63.1 & 7.1 & 2.7 & 0.31 & 0.28 & 0.04 & 0.29 & 0.16 & ~ &  & ~ &  \\ \hline 
\end{tabular}
\end{center}
\end{table*}

\section{Summary and conclusion}
\label{5}
Excitation functions of deuteron induced nuclear reactions on natural Yb were measured up to 50 MeV, as an extension and improvement of our earlier works. The comparison with the earlier experimental data, measured at lower energies, shows acceptable agreement (except for $^{173}$Lu and $^{175}$Yb).
The experimental data were compared with the results of our ALICE-D and EMPIRE-D calculations and with the data in the TENDL-2013 library based on TALYS 1.4 calculations. The theoretical descriptions of the experimental excitation functions in shape and in absolute values are acceptable if we consider the large disagreements of the earlier versions of the used model codes.
The obtained experimental data provide a basis for improved model calculations and for applications in different fields of nuclear medicine (e.g. cancer treatment), as tracer in nuclear biology, in industry and for non-destructive testing and as radioactive tracers in different processes (for related references see our earlier work \cite{2}).

\section{Acknowledgements}
\label{6}

This work was performed in the frame of the HAS-FWO Vlaanderen (Hungary-Belgium) project. The authors acknowledge the support of the research project and of the respective institutions in providing the beam time and experimental facilities.

\textbf{Correction}\\
During the EXFOR compilation of our previous paper on cross sections for deuteron induced processes on Yb (F. Tarkanyi et al., Nucl. Instrum. Meth. section B, 304(2013)36-48 \cite{2}) it was discovered that the decay data of $^{169}$Lu are misprinted in Table 1 of that publication. However, the cross section values presented in the tables and in the figures are correct. The $^{169}$Yb decay data used in the calculations are: T$_{1/2}$-34.04 h, E$_\gamma$(I$_\gamma$)-191.214 keV (20.6 \%) and 960.622 keV  (23.4 \%).	
 



\bibliographystyle{elsarticle-num}
\bibliography{Ybd}

\begin{thebibliography}{10}
\expandafter\ifx\csname url\endcsname\relax
  \def\url#1{\texttt{#1}}\fi
\expandafter\ifx\csname urlprefix\endcsname\relax\def\urlprefix{URL }\fi
\expandafter\ifx\csname href\endcsname\relax
  \def\href#1#2{#2} \def\path#1{#1}\fi

\bibitem{1}
F.~T\'ark\'anyi, A.~Hermanne, F.~Ditr\'oi, S.~Tak\'acs, B.~Kir\'aly, G.~Csikai,
  M.~Baba, H.~Yamazaki, M.~S. Uddin, A.~V. Ignatyuk, S.~M. Qaim, Systematic
  study of activation cross-sections of deuteron induced reactions used in
  accelerator applications (25-28 Oct., 2010 2011).

\bibitem{2}
F.~T\'ark\'anyi, F.~Ditr\'oi, S.~Tak\'acs, A.~Hermanne, H.~Yamazaki, M.~Baba,
  A.~Mohammadi, A.~V. Ignatyuk, Activation cross-sections of longer lived
  products of deuteron induced nuclear reactions on ytterbium up to 40 mev,
  Nuclear Instruments and Methods in Physics Research B 304 (2013) 36--48.

\bibitem{3}
A.~L. Nichols, R.~J. Bullock, P.~Glentworth, N.~R. Large, Excitation functions
  for the formation of various gadolinium and lutetium isotopes by (d,xn)
  reactions, Tech. rep., Atomic Energy Research Establishment, Harwell
  (England) (1972).

\bibitem{4}
A.~Hermanne, S.~Takacs, M.~B. Goldberg, E.~Lavie, Y.~N. Shubin, S.~Kovalev,
  Deuteron-induced reactions on yb: Measured cross sections and rationale for
  production pathways of carrier-free, medically relevant radionuclides,
  Nuclear Instruments and Methods in Physics Research Section B: Beam
  Interactions with Materials and Atoms 247~(2) (2006) 223--231.

\bibitem{5}
S.~Manenti, F.~Groppi, A.~Gandini, L.~Gini, K.~Abbas, U.~Holzwarth,
  F.~Simonelli, M.~Bonardi, Excitation function for deuteron induced nuclear
  reactions on natural ytterbium for production of high specific activity
  lu-177g in no-carrier-added form for metabolic radiotherapy, Applied
  Radiation and Isotopes 69~(1) (2011) 37--45.

\bibitem{6}
P.~P. Dmitriev, N.~N. Krasnov, G.~A. Molin, Radioactive nuclide yields for
  thick target at 22 mev deuterons energy, Yadernie Konstanti 34~(4) (1982) 38.

\bibitem{7}
S.~Manenti, M.~L. Bonardi, L.~Gini, F.~Groppi, Physical optimization of
  production by deuteron irradiation of high specific activity 177glu suitable
  for radioimmunotherapy, Nuclear Medicine and Biology in print~(0).

\bibitem{8}
F.~T\'ark\'anyi, S.~Tak\'acs, F.~Ditr\'oi, A.~Hermanne, H.~Yamazaki, M.~Baba,
  A.~Mohammadi, A.~V. Ignatyuk, Activation cross-sections of deuteron induced
  nuclear reactions on neodymium up to 50 mev, Nuclear Instruments and Methods
  in Physics Research Section B: Beam Interactions with Materials and Atoms
  325~(0) (2014) 15--26.

\bibitem{9}
F.~T\'ark\'anyi, S.~Tak\'acs, K.~Gul, A.~Hermanne, M.~G. Mustafa, M.~Nortier,
  P.~Oblozinsky, S.~M. Qaim, B.~Scholten, Y.~N. Shubin, Z.~Youxiang, Beam
  monitor reactions (chapter 4). charged particle cross-section database for
  medical radioisotope production: diagnostic radioisotopes and monitor
  reactions., Tech. rep., IAEA (2001).

\bibitem{10}
Canberra,
  http://www.canberra.com/products/radiochemistry\_lab\-/genie-2000-software.asp.
  (2000).

\bibitem{11}
G.~Sz\'ekely, Fgm - a flexible gamma-spectrum analysis program for a small
  computer, Computer Physics Communications 34~(3) (1985) 313--324.

\bibitem{12}
F.~T\'ark\'anyi, F.~Szelecs\'enyi, S.~Tak\'acs, Determination of effective
  bombarding energies and fluxes using improved stacked-foil technique, Acta
  Radiologica, Supplementum 376 (1991) 72.

\bibitem{13}
NuDat, Nudat2 database (2.6) http://www.nndc.bnl.gov/nudat2/ (2014).

\bibitem{14}
B.~Pritychenko, A.~Sonzogni, Q-value calculator, http://www.nndc.bnl.gov/qcalc
  (2003).

\bibitem{15}
H.~H. Andersen, J.~F. Ziegler, Hydrogen stopping powers and ranges in all
  elements. The stopping and ranges of ions in matter, Volume 3., The Stopping
  and ranges of ions in matter, Pergamon Press, New York, 1977.

\bibitem{16}
I.-B. of-Weights-and Measures, Guide to the expression of uncertainty in
  measurement, 1st Edition, International Organization for Standardization,
  Genève, Switzerland, 1993.

\bibitem{17}
A.~I. Dityuk, A.~Y. Konobeyev, V.~P. Lunev, Y.~N. Shubin, New version of the
  advanced computer code alice-ippe, Tech. rep., IAEA (1998).

\bibitem{18}
M.~Herman, R.~Capote, B.~V. Carlson, P.~Oblozinsky, M.~Sin, A.~Trkov,
  H.~Wienke, V.~Zerkin, Empire: Nuclear reaction model code system for data
  evaluation, Nuclear Data Sheets 108~(12) (2007) 2655--2715.

\bibitem{19}
F.~T\'ark\'anyi, A.~Hermanne, S.~Tak\'acs, F.~Ditr\'oi, I.~Spahn, S.~F.
  Kovalev, A.~V. Ignatyuk, S.~M. Qaim, Activation cross sections of the
  tm-169(d,2n) reaction for production of the therapeutic radionuclide yb-169,
  Applied Radiation and Isotopes 65~(6) (2007) 663--668.

\bibitem{20}
F.~T\'ark\'anyi, A.~Hermanne, S.~Tak\'acs, K.~Hilgers, S.~F. Kovalev, A.~V.
  Ignatyuk, S.~M. Qaim, Study of the 192os(d,2n) reaction for, production of
  the therapeutic radionuclide 192ir in no-carrier added form, Applied
  Radiation and Isotopes 65~(11) (2007) 1215--1220.

\bibitem{21}
A.~J. Koning, S.~Hilaire, M.~C. Duijvestijn, Talys-1.0 (2007).

\bibitem{22}
A.~J. Koning, D.~Rochman, S.~van~der Marck, J.~Kopecky, J.~C. Sublet, S.~Pomp,
  H.~Sjostrand, R.~Forrest, E.~Bauge, H.~Henriksson, O.~Cabellos, S.~Goriely,
  J.~Leppanen, H.~Leeb, A.~Plompen, R.~Mills, Tendl-2013: Talys-based evaluated
  nuclear data library (2012).

\end{thebibliography}







\end{document}